\documentclass[aps,prl,twocolumn,groupedaddress,floatfix]{revtex4}

\usepackage{graphicx}
\usepackage{dcolumn} 
\usepackage{bm}      

\begin{document}

\title{Contact Measurements on Atomic BEC}
\author{R. J. Wild, P. Makotyn, J. M. Pino, E. A. Cornell, and  D. S. Jin}

\affiliation{JILA, National Institute of
Standards and Technology and University of Colorado, and Department of Physics, Boulder, CO 80309-0440, USA}

\date{\today} \begin{abstract}
A powerful set of universal relations, centered on a quantity called the contact, connects the strength of short-range two-body correlations to the thermodynamics of a many-body system with delta-function interactions.
We report on measurements of the contact, using RF spectroscopy, for an $^{85}$Rb atomic Bose-Einstein condensate (BEC). For bosons, the fact that contact spectroscopy can be used to probe the gas on short timescales is useful given the decreasing stability of BECs with increasing interactions.  A complication is the added possibility, for bosons, of three-body interactions.  In investigating this issue, we have located an Efimov resonance for $^{85}$Rb atoms with loss measurements and thus determined the three-body interaction parameter.  In our contact spectroscopy, in a region of observable beyond-mean-field effects, we find no measurable contribution from three-body physics.
\end{abstract}

 \pacs{??)}

\maketitle


Systems with strong quantum correlations represent a frontier in our understanding of the complex quantum systems found in nature, and
atomic Bose-Einstein condensates (BEC) provide a versatile system in which to explore beyond mean-field physics.
Ultracold atoms experience two-body, short-range interactions that are well described theoretically by a delta-function pseudopotential characterized by an s-wave scattering length $a$.  In the simplest BEC experiments the values of $a$ and of the density $n$ are such that interactions are too weak, compared to the kinetic energy cost of correlations, to take the gas out of the mean-field regime.  The presence of a lattice potential can greatly suppress this kinetic energy cost, thus freeing the system to explore a much richer portion of many-body state space \cite{Morsch2006}.   The application of an external lattice potential, however, imposes an artificial orderliness not found in bosons in the wild. To explore strong interactions in a more naturalistic bulk three-dimensional gas, one can increase $a$ by means of a magnetic-field-tunable Feshbach scattering resonance \cite{Chin2010}.  Such efforts are motivated for instance by a desire to make better conceptual connections to the iconic strongly correlated fluid, liquid helium.

In practice it has proven difficult to study atomic BEC with increasing $a$ and only a few experiments have measured beyond-mean-field interaction effects in these systems  \cite{Papp2008a,Navon2010a,Smith2011}.  The difficulty comes from the fact that an increase in $a$ is accompanied by a dramatic increase in the rate of inelastic three-body processes \cite{Fedichev1996b,Esry1999b}.  This leads to large losses and significant heating of the trapped gas on a timescale similar to that for global equilibrium of the trapped cloud.  Probes of the gas that require global equilibrium, such as measurements of the density distribution or the amplitude or frequency of collective density oscillations in a trap, are therefore limited to systems that are only modestly out of the mean-field regime.  Our strategy for exploring BEC with larger interaction strengths is to start from an equilibrated weakly interacting gas, change the interaction strength relatively quickly, forsaking global equilibrium, and then use a fast probing technique to look at local many-body equilibrium in the trapped gas \cite{Papp2008a}. In this paper, we develop RF contact spectroscopy as a fast probe of short-range correlations in the BEC.

A central challenge in many-body physics lies in elucidating the dependence of an interacting many-body system on the strength of the few-body interactions.  For example, a fundamental theory result for interacting BECs is the energy density as a function of $a$ in the perturbative beyond-mean-field regime, first predicted by Lee, Huang, and Yang (LHY). For ultracold Fermi gases, it has been shown that the dependence of the energy on $a$ can be connected to the strength of two-particle short-range correlations through a set of universal relations that were introduced by Shina Tan  \cite{Tan2008a,Tan2008b,Tan2008c}.
These universal relations, which involve a quantity termed the ``contact", are extremely general, in that they hold true for any locally equilibrated gas regardless of the temperature, interaction strength, or number of particles. Tan's predictions have been explored theoretically \cite{Braaten2008a,Braaten2008b,Zhang2009a,Haussmann2009a,Blume2009,Werner2009a} and verified experimentally \cite{Kuhnle2010,Stewart2010a} for strongly interacting Fermi gasses.  The question we now address is whether contact spectroscopy can be used to probe interacting bosons.

The derivation of Tan's universal relations does not depend directly on the quantum statistics of the particles, however, it does assume that the interactions are fully described by a single parameter, $a$.  While this is true for an ultracold two-component (spin-up and spin-down) Fermi gas, it is in general not true for a Bose gas, where three-body interactions give rise to Efimov resonances \cite{Efimov1970}.   A number of recent experiments probing few-body physics in ultracold Bose gases have observed Efimov resonances \cite{Kraemer2006,Knoop2009,Zaccanti2009,Pollack2009,Gross2009}, however, many-body effects of the three-body interactions have not been observed.
To explore contact spectroscopy for bosons, we begin by examining RF spectroscopy assuming that three-body interactions do not significantly affect this measurement. Following this, we present a measurement of the three-body parameter for $^{85}$Rb using trap loss rates for a non-condensed gas, and look for many-body effects manifested in a three-body contact, $C_3$ \cite{Braaten2011,Castin2011}.

The two-body contact, $C_2$, is an extensive thermodynamic variable that is connected to the derivative of the total energy of the system, $E$, with respect to $a$ \cite{Tan2008b}. For bosons, the adiabatic sweep theorem states that \cite{Combescot2009a,Schakel2010a}
\begin{equation}\label{eq:AdiabaticSweep}
\frac{dE}{da}=\frac{\hbar^2}{8\pi m a^2}C_2.
\end{equation}
Combining this with the energy density of a BEC predicted by Lee, Huang, and Yang (LHY) \cite{Lee1957b},
 the predicted contact for a condensate is
\begin{equation}\label{eq:LHYcontact}
C_{2}=16\pi^2na^2\left(1+\frac{5}{2}\frac{128}{15\sqrt{\pi}}\sqrt{na^3}+...\right)N_{0}
,\end{equation}
where $n$ is the atom number density, $m$ is the atomic mass, and $N_0$ is the number of atoms in the BEC.

To measure $C_2$ using RF spectroscopy  \cite{Schneider2009a,Perali2008a}, an RF pulse drives a Zeeman transition and transfers a small fraction of spin-polarized bosonic atoms into another spin state, which we refer to as the final state. Interactions give rise to an asymmetric tail in the RF spectrum, which can be thought as RF ``dissociation" of pairs of atoms that happen to be very close to each other.  Ignoring $C_3$, and assuming that the measurement is done in the linear regime, the rate for transferring atoms to the final state in this tail is given by \cite{Braaten2010}
\begin{equation}\label{eq:RFContact}
\lim_{\omega \rightarrow
\infty} \Gamma(\omega) =   \frac{\Omega^2}{4\pi}\sqrt{\frac{\hbar}{m}} \frac{\alpha(a)}{\beta(\omega)}\frac{C_2}{\omega^{3/2}}
,\end{equation}
where the integrated RF lineshape is
$\int_{-\infty}^{\infty} \Gamma(\omega)d\omega=\pi\Omega^2 N$,
$\Omega$ is the Rabi frequency, and $N$ is the total number of atoms. In Eqn.\ \ref{eq:RFContact}, $\alpha(a)/\beta(\omega)$ describe final-state effects; the $a$-dependent part is $\alpha(a) = \left(a'/a-1\right)^2$, where $a'$ is the scattering length for interactions between atoms in the final spin state and atoms in the initial spin state, while the frequency-dependent part is $\beta(\omega) = 1+\hbar|\omega|/E'$, where $E'=\hbar^2/ma'^2$.

Our experiments probe 4-8$ \times10^4$ Bose-condensed $^{85}$Rb atoms in a gas with a 60$\%$ condensate fraction, and an average condensate density $\langle n\rangle$ of 4-10 x10$^{12}$ cm$^{-3}$. The atoms are in the $|F=2, m_F=-2\rangle$ state, where $F$ is the total atomic spin and $m_F$ is the spin projection. They are confined magnetically in a 10 Hz spherical harmonic trap with a variable magnetic bias field.  We work at magnetic-field values near a Feshbach resonance at 155.04 G \cite{Claussen2003a}, and during the final stages of evaporation, the field is set to give $a\sim$100 $a_0$. After evaporation, we ramp the bias field in order to change $a$ on a timescale that is fast compared to the trap period, but adiabatic with respect to two-body timescales, with $\dot{a}/a$ never reaching more than $0.01\hbar/(ma^2)$ ($\dot{a}$ being the time derivative of $a$) \cite{magfieldnote}.

\begin{figure}
\includegraphics[width=7 cm]{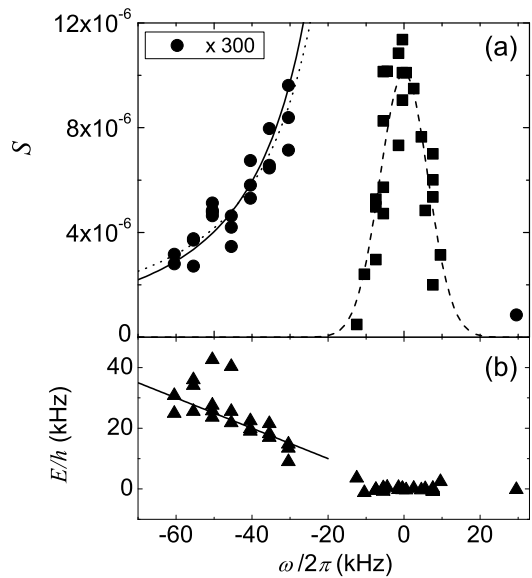}
\caption{Example of RF contact spectroscopy.
(a) RF lineshape, $S(\omega)$, normalized so that $\int_{-\infty}^{\infty}S(\omega)d\omega= 1\ \textrm{s}^{-1}$.
The data
at large detunings (circles) are multiplied by a factor of 300 to make the tail visible. The solid line is a fit to the expected frequency dependence from Eqn.\ \ref{eq:RFContact}, while the dotted line shows a fit ignoring $\beta(\omega)$. Here the mean density is $\langle n\rangle$ = 4.9 x10$^{12}$ cm$^{-3}$.
(b) Additional release energy of the outcoupled atom cloud. We calculate the energy from the width of the expanded cloud, $\sigma$, using $E=\frac{3}{2}m\frac{\sigma^2-\sigma_0^2}{\Delta t^2}$, where $\Delta t$ is the time between the middle of the RF pulse and the absorption image (4.5 ms) and $\sigma_0$ is the size of the expanded cloud measured at $\omega=0$. The solid line is $\frac{1}{2}\frac{|\omega|}{2\pi}$.}
\label{fig:LSandWidths}
\end{figure}

An example of RF contact spectroscopy at $a=497\pm5$ $a_0$, where $a_0$ is the Bohr radius, is shown in Fig.\ \ref{fig:LSandWidths}a. Roughly 1 ms after the magnetic-field ramps, we probe the BEC using a gaussian envelope RF pulse to drive the $|2, -2\rangle$ to $|2, -1\rangle$ transition. We determine $\Gamma(\omega)$ from the number of atoms transferred to the $|2, -1\rangle$ spin state divided by the RF pulse duration.  We then define our signal, $S(\omega)$, as $\Gamma(\omega)$ normalized by the integrated lineshape.  We fit $S(\omega)$ to a Gaussian lineshape (dashed black line in Fig.\ \ref{fig:LSandWidths}a) and take the center to be the single-particle transition frequency $\omega_0$. In general, the center of the RF lineshape will be shifted due to interactions, however we calculate the mean-field shift to be less than our typical fit uncertainty in $\omega_0/2\pi$ of $\pm0.5$ kHz. For the main lineshape, we use short RF pulses with a gaussian rms width for the field amplitude, $\tau$, of 5 $\mu$s; this sets the observed width of the lineshape.  At larger detunings, we use longer pulses, with an rms width of 25 to 200 $\mu$s, and an increased RF power, $\Omega^2$, such that we outcouple 1-2$\%$ of the gas.  We 
normalize the signal for the different $\tau$ and  $\Omega^2$, making small (5$\%$) corrections for measured nonlinearity in $\Omega^2\tau$.



\begin{figure}
\includegraphics[width=7 cm]{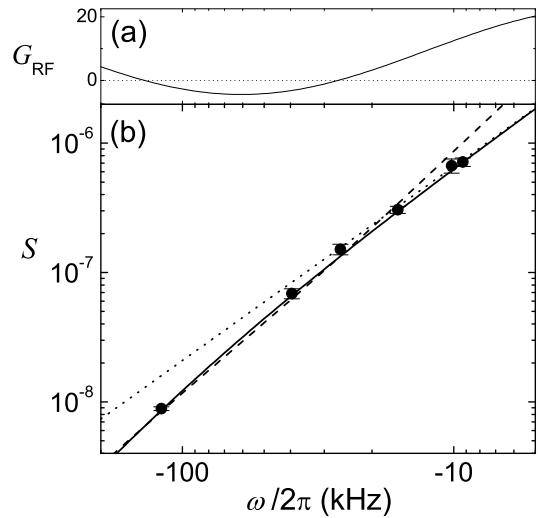}
\caption{(a) The frequency dependence of $G_\mathrm{RF}(\omega)$, given our measured value for $\kappa_*$. (b) Frequency dependence of the tail of the RF spectrum for $a=982\pm10$ $a_0$.  The solid line is a fit of the data ($\bullet$) to the expected frequency dependence of the two-body contact $C_2/N_0$ including final-state effects. The dotted line corresponds to the same value of $C_2/N_0$, but ignores final-state effects.  For comparison, the fit plus a trial $C_3/N_0$ term of 0.1 $\rm{\mu m^{-2}}$ is shown with the dashed line. Our measurements are consistent instead with a $C_3/N_0$ of zero. Here the mean density is $\langle n\rangle$ = 1.0 x10$^{13}$ cm$^{-3}$.}
\label{fig:logdata}
\end{figure}

For our experiment, the RF drives a transition to a lower energy spin state and one expects the $1/|\omega|^{3/2}$ interaction-induced tail on the low frequency side of the lineshape. Consistent with this expectation, we observe a tail for large negative detunings, while for similar detunings on the positive side, we find that the signal is consistent with zero.   The solid line in Fig.\ \ref{fig:LSandWidths}a shows a fit to the expected frequency dependence from Eqn.\ \ref{eq:RFContact}, while the dotted line shows a fit to $1/|\omega|^{3/2}$.  For our system, the final-state effects are characterized by $a'=-565$ $a_0$ \cite{Bohn2010a} and $E'/h=$133 kHz.  Over the range of the data shown here, the modification to the frequency dependence of the rf tail due to final-state effects is small. (Data for a wider range of $\omega$ is shown in Fig.\ \ref{fig:logdata}b.)


The $1/|\omega|^{3/2}$ tail, due to the contact, corresponds to an expected $1/k^4$ tail in the momentum distribution $n(k)$ \cite{Braaten2010,Stewart2010a}.  In Fig.\ \ref{fig:LSandWidths}b, we show the expansion energy of the outcoupled atoms, measured by releasing the gas from the trap and imaging the cloud after 3 ms of expansion.  In the region of the observed tail in the RF spectrum, the outcoupled atoms clearly have higher $k$ and our data show good agreement with the prediction (line in Fig.\ \ref{fig:LSandWidths}b) that the additional release energy should be $\frac{1}{2}\hbar|\omega|$, where the factor of $\frac{1}{2}$ comes from the assumption that the excess energy of the RF photon is shared between two pairwise interacting atoms \cite{Greiner2004}.

\begin{figure}
\includegraphics[width=7 cm]{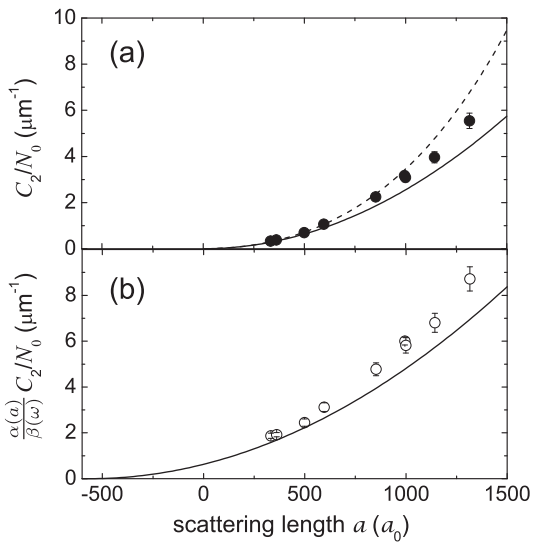}
\caption{The contact vs $a$, measured at $|\omega|=2\pi\times40$ kHz. (a) The contact per particle $\frac{C_2}{N_0}$. (b) The raw signal before final-state corrections. The solid lines in (a) and (b) show the mean-field predictions. The dashed line includes the next-order LHY correction. For increasing $a$, the data is consistently above the mean-field prediction.  For this data, the mean density is typically $\langle n\rangle$ = 5.8 x10$^{12}$ cm$^{-3}$, and we have linearly scaled the points to account for $\sim$10\% variation in density. The final-state effects shift the solid line from a parabola centered about $a=0$ in (a) to one centered about $a' = -565$ $a_0$ in (b), which enhances the raw signal at small $a$.}
\label{fig:signal}
\end{figure}



The strength of the RF tail, measured at $|\omega|=2\pi\times40$ kHz, is shown as a function of $a$ in Fig.\ \ref{fig:signal}.  As expected, we see the strength of the RF tail increase as $a$ increases.  In comparison with theory, our contact measurements are larger than the mean-field prediction (solid line in Fig.\ \ref{fig:signal}), but not as large as the prediction including the next order LHY term given in Eqn. \ref{eq:LHYcontact} (dashed line in Fig.\ \ref{fig:signal}). While beyond-mean-field physics is evident in the contact data shown here, we find that the measured strength of the RF tail depends on the speed of the magnetic-field ramp to increase $a$, with $C_2/N_0$ increasing for slower ramps. It will be important to carefully explore this intriguing dependence on ramp speed in order to make a quantitative comparison between the experiment and theory. Moreover, the fact that these ramps are still short compared to the timescale required for global equilibrium opens the exciting possibility for using RF contact spectroscopy to probe local dynamics in the beyond-mean-field regime.

We now turn our attention to $C_3$, which is connected to the derivative of $E$ with respect to a three-body interaction parameter $\kappa_*$ \cite{Braaten2011,Castin2011}
\begin{equation}\label{eq:AdiabaticSweep}
\frac{dE}{d \kappa_*}=-\frac{2\hbar^2}{m \kappa_*}C_3
.\end{equation}
Three-body short-range correlations contribute a predicted additional term to the RF tail at large detunings that should be added to the right-hand side of Eqn.\ \ref{eq:RFContact} \cite{Braaten2011}:
\begin{equation}\label{eq:C3term}
\frac{\hbar\Omega^2}{2 m} \frac{G_\mathrm{RF}(\omega)}{\omega^2}C_3
.\end{equation}
Here, $G_\mathrm{RF}(\omega)$ is a log-periodic function rooted in Efimov physics:
\begin{equation}\label{eq:LogPeriodic}
G_\mathrm{RF}(\omega)=9.23 -13.6\sin[s_0\ln(m|\omega|/\hbar\kappa_*^2)+2.66]
.\end{equation} Efimov physics predicts an infinite series of successively more weakly bound trimers whose binding energies at unitarity ($a\rightarrow\infty$) are given by $\frac{\hbar^2\kappa_*^2}{m}(e^{-2\pi/s_0})^l$, where $l$ is an integer and
 $s_0$ is 1.00624 for identical bosons \cite{Braaten2006}.   We note that there is as yet no prediction for final-state effects on the $C_3$ contribution to the RF tail.

 In order to determine $\kappa_*$ for $^{85}$Rb atoms, we have performed measurements of trap loss rates in a low temperature, non-condensed gas as a function of $a$.  With these measurements, we locate an Efimov resonance, which is a peak in the three-body recombination rate that occurs when the trimer energy becomes degenerate with the threshold for three unbound atoms.  Similar measurements of Efimov resonances have been reported for several other ultracold atom systems \cite{Kraemer2006,Knoop2009,Zaccanti2009,Pollack2009,Gross2009,Ottenstein2008}.  The value of $a$ for the resonance, $a_-$, is related to the three-body parameter through $\kappa_*=-1.56(5)/a_-$ \cite{Braaten2006}.

\begin{figure}
\includegraphics[width=7 cm]{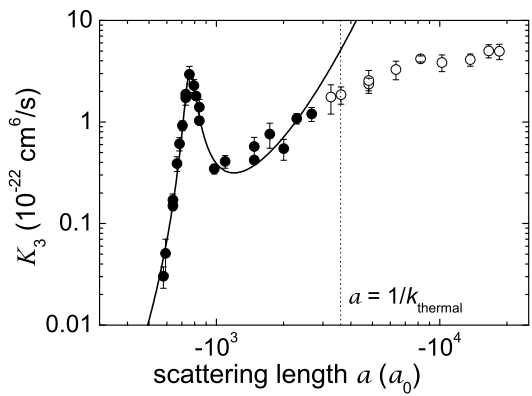}
\caption{A three-body loss resonance for $^{85}$Rb.  We plot the three-body event constant $K_3$ vs $a$.  From fitting Eqn.\ \ref{eq:Efimov} to the solid points, for which $a<1/k_\mathrm{thermal}$, we extract $a_-=-759(6) a_0$ and $\eta=0.057(2)$.}
\label{fig:efimov}
\end{figure}

The observed $^{85}$Rb Efimov resonance is shown in Fig.\ \ref{fig:efimov}.  For these measurements, we make non-condensed clouds of $1.5\times10^5$ atoms at a temperature $T=80$ nK. After ramping the magnetic field to realize the desired $a$ on the $a<0$ side of the Feshbach resonance,  we use absorption imaging to measure the number of atoms and cloud size as a function of hold time.  We then extract the three-body event rate constant $K_3$, which is defined by $\frac{d}{dt}N=-3 K_3 \langle n^2 \rangle N$ when all three atoms are lost per event. In extracting $K_3$, we assume that all of the measured loss is due to three-body processes and we account for the observed heating of the gas, which causes additional decrease in $n$ in time.  Our $500$ s vacuum-limited lifetime and previous experiments on $^{85}$Rb suggest that one- and two-body losses can be ignored for this range of magnetic fields \cite{Roberts2000a}. We fit the measured $K_3$ vs $a$ to the expected form for an Efimov resonance for non-condensed atoms \cite{Braaten2006},
\begin{equation}\label{eq:Efimov}
K_3 = \frac{4590 \sinh(2\eta)}{\sin^2[s_0\ln(a/a_-)]+\sinh^2\eta}\frac{\hbar a^4}{m}
.\end{equation}
Because this expression comes from a $T=0$ theory, we only fit the data for $a<1/k_\mathrm{thermal}$, where $k_\mathrm{thermal}=\sqrt{2mk_BT}/\hbar$ and $k_B$ is Boltzmann's constant.  From the fit, we extract $a_-=-759(6)$ $a_0$ and $\eta=0.057(2)$. This gives $\kappa_*$=39(1) $\rm{\mu m
}^{-1}$.

To see how the three-body parameter might impact the many-body physics, we plot the expected frequency dependence of $G_\mathrm{RF}(\omega)$  in Fig.\ \ref{fig:logdata}a.  Note that $G_\mathrm{RF}(\omega)$ has a node at $|\omega|\sim2\pi\times27$ kHz and a smaller magnitude at larger $|\omega|$.  Eqn.\
\ref{eq:C3term} has a frequency dependence given by $G_\mathrm{RF}(\omega)/\omega^2$, which suggests that the largest contribution from $C_3$ will be for smaller $|\omega|$.  The prediction for the $C_3$ term (Eqn.\ \ref{eq:C3term}), like the $C_2$ term (Eqn.\ \ref{eq:RFContact}), is valid for $\omega\rightarrow \infty$.  For the case of the $C_2$ term, the RF tail arises from two-body short-range correlations at distances that are small compared to the interparticle spacing, which requires $\omega\gg\hbar n^{2/3}/m$. For our typical experimental parameters, $\hbar n^{2/3}/m\sim1$ kHz and this requirement is always satisfied.  However, for the case of $C_3$, the prediction for the $C_3$  tail contribution to the RF tail may have a more limited range of applicability.  In particular, the $C_3$ theory may only be applicable for $|\omega|>\frac{\hbar}{ma^2}$ \cite{Braaten}, where the frequency dependence makes it less likely to contribute significantly to the RF tail.

The results of our search for $C_3$ can be seen in Fig.\ \ref{fig:logdata}b, where we examine the frequency dependence of the RF tail for a BEC at $a=982\pm10$ $a_0$.  Residual magnetic-field gradients broaden the central feature in the RF spectrum, and this limits our data for the tail to $|\omega|\geq2\pi\times 10$ kHz. In this frequency regime, we verify that technical contributions to the signal are negligible by checking that we detect no signal for positive detunings. We fit the data to the predicted frequency dependence of the $C_2$ contribution, shown by the solid line.  The dotted line is the same fit but shown without including the final-state correction $1/\beta(\omega)$.  We can see that our data fit very well to the expected frequency-dependence for the two-body contact with final-state effects, and we do not observe any deviation consistent with a three-body term.
Fitting the data to both contributions gives an upper limit for $C_3/N_0$ of  0.07 $\rm{\mu m^{-2}}$.

In the regime of perturbative interactions, such as assumed in the LHY calculation, one would expect that the short-range correlations in the BEC are dominated by two-body effects.  This is consistent with our measurements, where no clear signature of three-body effects is seen in the frequency dependence of the interaction-induced tail in RF spectroscopy.  In general, this paves the way for using RF spectroscopy to measure the two-body contact for BECs and thus measure beyond-mean-field physics and probe non-equilibrium many-body dynamics.

Moreover, three-body physics is itself very intriguing, and a result of our studies is the location of the $^{85}$Rb Efimov resonance. When $a_-$ is expressed in units of the mean scattering length of the van der Waals potential \cite{Gribakin1993a} for $^{85}$Rb (78.5 $a_0$), we find a value of -9.67(7) \cite{vanderWaalsnote}, which is very similar to reported results for $^{133}$Cs (for multiple Feshbach resonances) \cite{Berninger2011} and for $^7$Li \cite{Gross2009}.  This adds to the empirical evidence suggesting that the three-body parameter depends only on the coefficient of the $1/r^6$ part of the two-body potential and not on the details of a three-body potential at short range \cite{Berninger2011}.  In the many-body physics of an interacting BEC, three-body correlations may yet play a significant role outside of the regime relevant to the usual perturbative theoretical treatment. For example, it will be interesting to look for three-body effects on BECs with strong interactions (at unitarity), or at $a=a_-$. The techniques developed here, namely the investigation of the density dependence and frequency dependence of the RF tail, can be used to distinguish two-body and three-body interaction contributions to the many-body physics.

We acknowledge useful discussions with Eric Braaten, Paul Julienne, Jeremy Hutson, Shina Tan, John Bohn, and the larger Cornell/Jin groups. We thank Chris Poulton and Sophie Letournel for experimental assistance. This work is supported by the NSF, ONR, and NIST.

\bibliographystyle{prsty} 
\bibliography{Aug_2011}

\begin{thebibliography}{10}
\bibitem{magfieldnote}
Simultaneous with this ramp, we quickly shift the vertical position of the
  harmonic trap to minimize the magnetic-field variation across the cloud by
  overlapping the center of the magnetic-field curvature with the vertical
  position of the cloud (this was previously not the case because of
  gravitational sag). In this configuration, non-uniformity of the magnetic
  field contributes a 3-5 kHz spread in the single-particle RF resonance
  frequency.
\end{thebibliography}

\end{document}